\documentclass[prb,aps,showpacs,twocolumn,preprintnumbers,amsmath,amssymb]{revtex4}
\usepackage{graphicx}
\usepackage{dcolumn}
\usepackage{bm}

\begin{document}

\title{Spin Pumping of Current in Non-Uniform Conducting Magnets}
\author{ Wayne M. Saslow} 
\email{wsaslow@tamu.edu}
\affiliation{ Department of Physics, Texas A\&M University, College Station, TX 77843-4242}
\date{\today}

\begin{abstract}
Using irreversible thermodynamics we show that current-induced spin transfer torque within a magnetic domain implies spin pumping of current within that domain.  This has experimental implications for samples both with conducting leads and that are electrically isolated.  These results are obtained by deriving the dynamical equations for two models of non-uniform conducting magnets: (1) a generic conducting magnet, with net conduction electron density $n$ and net magnetization $\vec{M}$; and (2) a two-band magnet, with up and down spins each providing conduction and magnetism.  For both models, in regions where the equilibrium magnetization is non-uniform, voltage gradients can drive adiabatic and non-adiabatic bulk spin torques.  Onsager relations then ensure that magnetic torques likewise drive adiabatic and non-adiabatic currents -- what we call bulk spin pumping.  For a given amount of adiabatic and non-adiabatic spin torque, the two models yield similar but distinct results for the bulk spin pumping, thus distinguishing the two models.  As in the recent spin-Berry phase study by Barnes and Maekawa, we find that within a domain wall the ratio of the effective emf to the magnetic field is approximately given by $P(2\mu_{B}/e)$, where $P$ is the spin polarization.  The adiabatic spin torque and spin pumping terms are shown to be dissipative in nature. 

\end{abstract}

\pacs{72.25.-b, 75.60.Ch, 05.70.Ln}

\maketitle
\section{Introduction}
\subsection{Current-induced Spin Transfer Torque}
Current-induced spin transfer torque at both surfaces\cite{Berger96, Slonc96} and in bulk is by now a well-established phenomenon.\cite{STreview}  In surface spin transfer torque, when a polarized spin current from a non-magnet crosses the interface with a magnet it causes spin motion.  In bulk spin transfer torque, when a polarized spin current crosses a domain wall (where the magnetization varies in direction) it causes spin motion.  For a uniform magnet, there is no current-induced spin torque.

To be specific, we consider that the magnetization $\vec{M}$ satisfies the equation\cite{Baz98}
\begin{equation}
\partial_{t}\vec{M}+\partial_{i}\vec{Q}_{i}=-\gamma\vec{M}\times\vec{H}'+\vec{N},
\end{equation}
where $\vec{Q}_{i}$ is the magnetization flux (or magnetization current density), $-\gamma\vec{M}\times\vec{H}'$ is the Larmor-like torque, and $\vec{N}$ is the rest of the torque density acting on the magnetization.  We take $\gamma$ to be the magnitude of the (negative) gyromagnetic ratio.  

$\vec{H}'$ is the net effective field, which can include an external field, crystalline anisotropy, the demagnetization field, and non-uniform exchange.  (We will reserve $\vec{H}$ for a somewhat different quantity, which not only gives $\vec{M}\times\vec{H}=\vec{0}$ in equilibrium, but in fact satisfies $\vec{H}=\vec{0}$ in equilibrium.) 
$\vec{H}'$ is given in SI units of T, although $\vec{H}'$ is not the magnetic induction field $\vec{B}$, whose units also are in T; an H-field in A/m, on multiplication by $\mu_{0}$, becomes an H-field in T.  

We will call 
\begin{equation}
\vec{N}'=\vec{N}-\partial_{i}\vec{Q}_{i}
\label{N'}
\end{equation}
the non-Larmor-like spin torque.  (The prime here is unrelated to the prime on $\vec{H}'$.)  When $\vec{N}'$ contains a term that is proportional to the current (or to the gradient of the electrochemical potential) one says that there is a spin transfer torque.  

For a uniform system, $\vec{N}'$ will contain only damping terms.  However, for a nonuniform system $\vec{N}'$ also contains terms of the form\cite{ZhangLi04}
\begin{equation}
-\xi\partial_{i}\tilde\mu[\partial_{i}\vec{M}-\beta\partial_{i}\hat{M}\times\partial_{i}\vec{M}], 
\end{equation}
where $\tilde\mu$ the electrochemical potential, $-\xi\partial_{i}\tilde\mu$ has units of a velocity, and $\beta$ is dimensionless.  The first term is called an adiabatic spin transfer torque and the second term is called a nonadiabatic spin transfer torque.\cite{ad-nonad}  

In the literature the current $j_{i}$ usually is written (with a suitable conductivity) in place of $\partial_{i}\tilde\mu$, but when one employs irreversible thermodynamics $\partial_{i}\tilde\mu$, a thermodynamic driving force that is even under time-reversal, is the more natural quantity to employ.  Use of $\partial_{i}\tilde\mu$ unambiguously leads to the conclusion that, because of their time-reversal properties, the adiabatic spin transfer torque is irreversible (dissipative) and the nonadiabatic spin transfer torque is reversible (reactive).  This also will be seen from a calculation of the volume rate of dissipation $R$, where the equivalent of $\xi$ appears, but the equivalent of $\beta$ is absent.  

\subsection{Spin Pumping of Current}
Spin pumping of current at surfaces is also a well-established phenomenon.\cite{TBBH-RMP}  Here spin dynamics at an interface transfers a spin-polarized current to an adjacent material.  The first indication of spin pumping was provided by experiments on a thin magnetic film adjacent to both a vacuum and an ordinary conductor.\cite{Monod72}  The present work studies two models for non-uniform conducting magnets, one a generic conducting magnet that is related to but distinct from the s-d model,\cite{ZhangLi04} and one based on a two-band magnet.\cite{SXZ04}

For both models, Onsager relations between transport coefficients imply a bulk version of spin pumping, to an extent related to the amount of bulk spin transfer torque.  (This is analogous to how, if temperature gradients can cause an electric current, then electrochemical potential gradients can cause a heat current, with the size of these effects related by Onsager relations.)  However, the effects are somewhat different for the two models, permitting them to be distinguished.  

For the band model, there are two currents (from up and down spins) and two effective electrochemical potentials.  Each of these currents can be ``spin-pumped'' by disequilibrium of $\vec{M}$.  For the generic conducting magnet, there is only one current, but it is taken to be spin-polarized.  For the generic conducting magnet we study the number current density $j^{n}_{i}$ [cf. (\ref{jn-sd})].  We interpret those terms with the form 
\begin{equation}
j^{n}_{i}=-L_{nM1}(\partial_{i}\vec{M})\cdot\vec{H}-L'_{nM}\partial_{i}\vec{M}\cdot(\vec{M}\times\vec{H})+\dots
\end{equation}
as representing the adiabatic and non-adiabatic spin pumping terms.  Bulk spin-pumping of current requires a non-uniform magnetization; it occurs within domains.  

Barnes and Maekawa have recently studied the effective electric field associated with the spin-Berry phase induced by the domain structure.\cite{BarnesMaekawa07}  The present results are remarkably similar to their, despite the vastly different methods.  In particular, both works find that the ratio of the effective emf to the magnetic field is given by $P(2\mu_{B}/e)$, where $P$ is the spin polarization, although our result does not appear to be exact. 

The spin pumping terms are related to the spin transfer torque terms by Onsager relations.  (In surface spin pumping, there is also a spin transfer torque in proportion to the spin pumping.\cite{TBBH-RMP}) Thus, not only does the fact that spin transfer torque has been observed tell us that spin pumping must exist (if we understand the correct thermodynamic description of our system), it also tells us how large the spin pumping terms must be.  For the two-band magnet, analogous terms appear in each of the two currents.   However, the two models have forms that permit them to be distinguished.  Thus the theory may also provide a means to distinguish between the two models.\cite{s-d}  

\subsection{Summary of Methods}
This paper employs the methods of irreversible thermodynamics.  For an introduction to these methods that is directed to the magnetism community, see Ref.\onlinecite{RivkinSaslow}.  We summarize the approach as follows.\cite{Forster, Reichl, ChaikinLubensky} 

$\bullet$The first step is to determine the appropriate thermodynamic variables and the thermodynamic relation for the differential $d\varepsilon$ of the energy density as a sum of terms proportional to the various thermodynamic densities describing the system.  For an ordinary one-component system these densities are associated with the entropy and the particle number.  For the generic magnetic conductor, these densities are the entropy density $s$, the density of carrier electrons $n$, and the magnetization $\vec{M}$.  For a two-component non-magnetic system like a semiconductor these densities are associated with the entropy and the particle numbers (electrons and holes).\cite{KrcmarSas}  For the two-band magnet, these densities are associated with the entropy and the particle numbers (up and down electrons, with particle densities $n_{\uparrow}, n_{\downarrow}$, and the magnetization direction $\hat{M}$. 

$\bullet$The second step is to require that all of the densities satisfy either a conservation law (with unknown fluxes) or a source equation (with unknown sources) or both.  

$\bullet$The third step is to construct the matrix of structure-dependent constants (the transport and dissipation coefficients) relating the sources and fluxes to the thermodynamic forces.  Essential to the construction of this matrix are the time-reversal properties of the fluxes, forces, and sources.  Finally, the Onsager relations are used to reduce the number of independent coefficients.

The structure of this paper will be, in each section, to develop the appropriate parts of each theory in parallel.  

\begin{figure}
  \vskip-2.5in
  \includegraphics[angle=0,width=4.5in]{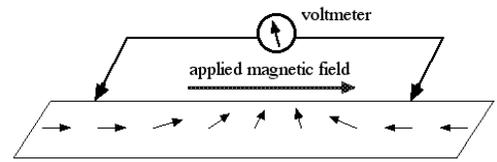}
  \vskip-2.5in
  \caption{Experimental geometry to observe spin-pumping of current associated with a head-to-head domain wall.  An applied magnetic field drives the domain wall to the right.  A spin-pumped current goes to the left.  One expects an associated voltage pulse when the domain wall crosses either voltage lead.}
  \label{fig1}
  \vskip-0.1in
\end{figure}

\subsection{Experimental Implications}

The most important prediction of this work, based on irreversible thermodynamics, is that bulk spin pumping occurs if there is spin transfer torque (and vice-versa); moreover, the amount of spin pumping can be determined from appropriate Onsager relations.  

For a magnetic wire with head-to-head domains, and two voltage leads, if a domain wall is $\vec{H}$-field-driven past one lead there should be a voltage jump.  Figure 1 shows an experimental geometry corresponding to a linear head-to-head domain wall that is driven by an external field.  When the domain wall crosses the voltage lead on the right, the voltage difference between the leads will measure a pulse that drives current leftward.  Likewise, for a magnetic dot with a vortex configuration, and two voltage leads, if the vortex structure is $\vec{H}$-field-driven past one lead there should be a voltage jump. 

For Co, data on spin torque indicates that within a domain wall a true $E$-field $E_{0}=1.0\times10^{4}$~V/m can cause the same spin torque as direct application of a magnetic field $H_{ST}=4\times10^{-3}$~T.   From this, 
in Sect.~VII we estimate that within a domain wall a true $H$-field $H_{0}=0.1$~T can cause the same spin-pumped current as direct application of an electric field $E_{SP}=350$~(V/m), or, across the domain wall, can cause the same effect as a voltage difference of $7.0\times10^{-6}$~V.  In terms of the effective emf per magnetic field, given by $P(2\mu_{B}/e)$, this is $0.7\times10^{-4}$~V/T.






\section{Energy and Entropy}
Energy and entropy are treated the same way in both theories. 

Consider the energy density $\varepsilon$, which, being conserved, has only a flux:  
\begin{equation}
\partial_{t}\varepsilon+\partial_{i}j^{\varepsilon}_{i}=0,
\label{de/dt}
\end{equation}
Here $j^{\varepsilon}_{i}$ is the energy flux density.  The intrinsic time-reversal signature of $\varepsilon$ is even, so the intrinsic time-reversal signature of $j^{\varepsilon}_{i}$ is odd.  

Now consider the entropy density $s$.  It has both a flux $j^{s}_{i}$ and a source $R/T$, where $R$ is the volume rate of heat production: 
\begin{equation}
\partial_{t}s+\partial_{i}j^{s}_{i}=\frac{R}{T}\ge0.
\label{ds/dt}
\end{equation}
Irreversible thermodynamics considers the time-behavior of thermodynamic variables, which have definite signature under time-reversal.  As a consequence, in an equation of motion for that quantity that has both a flux and a source, because of the time-derivative, the intrinsic time-reversal signatures of both the flux (here $j^{s}_{i}$) and the source (here $R/T$) are opposite to the intrinsic time-reversal signature of then extensive density.  Since $s$ is even, the intrinsic time-reversal signature of $j^{s}_{i}$ and $R/T$ are odd.  

\section{Thermodynamic relations}

\subsection{Generic Conducting Magnet}
We consider a system with magnetization $\vec{M}$ that is basically due to localized electrons, of gyromagnetic ratio $-\gamma$, and conduction electrons with electrochemical potential $\tilde\mu$ and number density $n$.  We take the thermodynamics to be given by
\begin{equation}
d\varepsilon=Tds+\tilde\mu dn-\vec{H}\cdot d\vec{M}.
\label{de-sd}
\end{equation}
Here $\vec{H}$ includes not only the field $\vec{H}'$ that causes the Larmor torque, but also a uniform 
exchange field that points along $\vec{M}$, and thus has no effect on the Larmor torque.  The exchange field is chose to make $\vec{H}_{eq}=\vec{0}$ in equilibrium, so that $\vec{H}$ will serve as a thermodynamic force.  

To be specific, consider a uniform ferromagnet in a minor loop and in an external field $\vec{H}_{0}$.  In the absence of anisotropy, in equilibrium it satisfies $\vec{M}=M_{0}\hat{H}_{0}+\chi\vec{H}_{0}$.  Let there also be an anisotropy field $\vec{H}_{\rm an}$.  Further, include an a non-uniform exchange term $A'(\partial_{i}\vec{M})^{2}$ in the energy density, which will yield a non-uniform exchange field $\vec{H}_{\rm ex}=2A'\nabla^{2}\vec{M}$.  We assert that 
\begin{eqnarray}
\vec{H}&=&-\frac{\delta\varepsilon}{\delta\vec{M}}=-\frac{\partial\varepsilon}{\partial\vec{M}}+\partial_{i}\frac{\partial\varepsilon}{\partial(\partial_{i}\vec{M})}\cr
&=&[\vec{H}_{0}+\vec{H}_{\rm an}+2A'\nabla^{2}\vec{M}]-[\frac{\vec{M}-M_{0}\hat{M}}{\chi}]\cr
&\equiv&\vec{H}'-\vec{H}_{int}.
\label{Hf}
\end{eqnarray}
has the desired form.  The first bracket, with three terms, constitutes $\vec{H}'$, which drives Larmor precession.  The last bracket, with two terms, represents an internal field $\vec{H}_{int}$ due to exchange.  In equilibrium, $\vec{H}_{int}=\vec{H}'$, to yield $\vec{H}=\vec{0}$ in equilibrium. 

Setting $\vec{H}=\vec{0}$ gives $\vec{M}$ along $\vec{H}'$, as desired for no Larmor torque.  Moreover, it gives $\vec{M}=M_{0}\hat{H}'+\chi\vec{H}'$, as expected.    Even out of equilibrium $\vec{H}_{int}$, which is along $\hat{M}$, does not contribute to $\vec{M}\times\vec{H}$, so that $\vec{M}\times\vec{H}'=\vec{M}\times\vec{H}$, a substitution we will make when needed.  In the absence of anisotropy and non-uniform exchange, (\ref{Hf}) very similar to the form employed by Ref.\onlinecite{JohnsonSilsbee87}, but called $\vec{H}^{*}$. 


In the s-d model as often used, there is an exchange field that couples to the $s$ electrons, giving them a weak magnetization.  If one wants to incorporate this idea in the present framework, then one may consider that $\vec{M}$ includes a contribution from the polarized conduction electrons.  However, because no such specification is made, our generic conducting magnet is distinct from the s-d model. 

\subsection{Two-band Magnet}
The two-band magnet\cite{SXZ04} considers a conducting magnetic system consisting of electrons of charge $-e$ and gyromagnetic ratio $-\gamma$, where $\gamma=|g|\mu_{B}/\hbar>0$ and $\mu_{B}=e\hbar/2m$ (for free electrons we take $g=-2$).  We assume that the electrons partially occupy two, spin-dependent, conduction bands, with number densities $n_{\uparrow}$ and $n_{\downarrow}$.  For specificity we assume that $n_{\downarrow}>n_{\uparrow}$, so that the magnetization will point along the ``up'' direction, determined either by spontaneous symmetry breaking or by an external field and anisotropy.  The system also has an entropy density $s$.  The total number density for the conducting electrons is 
\begin{equation}
n=n_{\uparrow}+n_{\downarrow}.
\label{n}
\end{equation}
Moreover, the magnetization is given by 
\begin{equation}
\vec{M}=-\gamma\vec{S},
\label{vecM0}
\end{equation}
where $\vec{S}$ is the spin density, of magnitude 
\begin{equation}
S=(\hbar/2)(n_{\downarrow}-n_{\uparrow}).
\label{decayrate}  
\end{equation}
Thus the magnetization has magnitude $M=|\vec{M}|$ given by
\begin{equation}
M=\gamma(\hbar/2)(n_{\downarrow}-n_{\uparrow})=(|g|\mu_{B}/2)(n_{\downarrow}-n_{\uparrow}).
\label{magM}
\end{equation}  
With magnetization direction $\hat{M}$, we then have 
\begin{equation}
\vec{M}=\gamma(\hbar/2)(n_{\downarrow}-n_{\uparrow})\hat{M}.
\label{vecM}
\end{equation}
Note that  
\begin{equation}
d\vec{M}=\gamma(\hbar/2)\hat{M}\,d(n_{\downarrow}-n_{\uparrow})+M\,d\hat{M}.
\label{dvecM}
\end{equation}
For the differential of the energy density we take \begin{equation}
d\varepsilon=Tds+\mu_{\uparrow}^{*}dn_{\uparrow}+\mu_{\downarrow}^{*}dn_{\downarrow}-M(\vec{H}\cdot d\hat{M}).
\label{de-2b}
\end{equation}
Here 
\begin{equation}
\mu^{*}_{\uparrow,\downarrow}=\tilde\mu_{\uparrow,\downarrow}\pm\gamma(\hbar/2)(\vec{H}\cdot\hat{M}),
\label{mu*-2b}
\end{equation}
where
\begin{equation}
\tilde\mu_{\uparrow, \downarrow}=\mu_{\uparrow, \downarrow}-eV
\label{mu-2b}
\end{equation}
is the electrochemical potential in terms of the chemical potential, with $V$ being the electrical potential.  
Changes in the number density are affected both by the electrochemical potential and $\vec{H}$.  In equilibrium $\mu_{\uparrow}^{*}=\mu_{\downarrow}^{*}$ and $\hat{M}\times\vec{H}=\vec{0}$.   If we further require that $\vec{H}=\vec{0}$ in equilibrium, then $\tilde\mu_{\uparrow}=\tilde\mu_{\downarrow}$ in equilibrium. 

\section{Conservation Laws and Equations of Motion}
\subsection{Generic Conducting Magnet}
We take the equations of motion and conservation laws for the four variables 
$n$ and $\vec{M}$ to be: 
\begin{eqnarray}
\partial_{t}n+\partial_{i}j^{n}_{i}&=&0,\label{dn/dt-sd}\\
\partial_{t}\vec{M}+\partial_{i}\vec{Q}_{i}&=&-\gamma\vec{M}\times\vec{H}+\vec{N}.\label{dvecM/dt-sd}
\end{eqnarray}
Here $j^{n}_{i}$ is the number current density, $\vec{Q}_{i}$ is the magnetization flux density ($i$ is the real space index), and $\vec{N}$ (a source) is the volume rate of change of magnetization due to torques associated with a lack of thermal equilibrium.  We have employed $\vec{M}\times\vec{H}'=\vec{M}\times\vec{H}$, thus substituting $\vec{H}$, which in equilibrium is zero, for the precession-causing field $\vec{H}'$.\cite{Q-def}

\subsection{Two-band Magnet}
We take the equations of motion and conservation laws for the four variables 
$n_{\uparrow}$, $n_{\downarrow}$, and $\hat{M}$ to be: 
\begin{eqnarray}
\partial_{t}n_{\uparrow}+\partial_{i}j_{\uparrow i}&=&S,\label{dnup/dt}\\
\partial_{t}n_{\downarrow}+\partial_{i}j_{\downarrow i}&=&-S,\label{dndown/dt}\\
\partial_{t}\hat{M}&=&(\gamma \vec{H}+\vec{\Omega})\times\hat{M}\label{dhatM/dt}.
\end{eqnarray}
Here $j_{{\uparrow},i}$ and $j_{{\downarrow},i}$ are the number current densities, $S$ is the decay rate for up spins (by charge conservation this is compensated by the decay rate $-S$ for down spins), and $\gamma \vec{H}$ and $\vec{\Omega}$ are the Larmor and non-Larmor parts of the rotation rate for $\hat{M}$.  By definition $\vec{\Omega}$ has only two components, and is normal to $\vec{M}$.  (Again we have made the allowable substitution, in the torque, of $\vec{H}$ for $\vec{H}'$.) 

\subsection{Two-band Magnet: Implied Equations}
The above equations imply certain equations of motion for $n$ of (\ref{n}), $M$ of (\ref{magM}) and $\vec{M}$ of (\ref{vecM}).  
These equations are not necessary, because the previous section is self-contained, but they are useful for comparison with previous work.

{\bf Continuity equation.}
Eqs. (\ref{n}), (\ref{dnup/dt}), and (\ref{dndown/dt}) imply that
\begin{equation}
\partial_{t}n+\partial_{i}j^{n}_{i}=0, \qquad j^{n}_{i}\equiv j_{\uparrow i}+j_{\downarrow i} 
\label{dn/dt-2b}.
\end{equation}
With the charge density $\rho=-en$ and current density (charge flux) given by 
\begin{equation}
\rho=-en, \qquad j_{i}=-ej^{n}_{i}=-e(j_{\uparrow i}+j_{\downarrow i}),
\label{rhoJ-2b}
\end{equation}
the continuity equation is automatically satisfied: 
\begin{equation}
\partial_{t}\rho+\partial_{i}j_{i}=0.
\label{drho/dt}
\end{equation}

{\bf Magnitude of Magnetization.}
Eqs. (\ref{magM}), (\ref{dnup/dt}), and (\ref{dndown/dt}) imply that
\begin{equation}
\partial_{t}M+\gamma(\hbar/2)\partial_{i}(j_{\downarrow i}-j_{\uparrow i})=-2S\gamma(\hbar/2).
\label{dmagM/dt}
\end{equation}
For magnetization along $z$, this is analogous to $\partial_{t}M_{z}$. 

{\bf Vector Magnetization.}
Eqs. (\ref{vecM}), (\ref{dnup/dt}), (\ref{dndown/dt}), and (\ref{dhatM/dt}) imply that 
\begin{equation}
\partial_{t}\vec{M}+\gamma(\hbar/2)\hat{M}\partial_{i}(j_{\downarrow i}-j_{\uparrow i})=-\gamma\vec
{M}\times\vec{H}+\vec{\Omega}\times\vec{M}-2S\gamma(\hbar/2)\hat{M}.
\label{dvecM/dt-2b}
\end{equation}
This can be rewritten in the more conventional form
\begin{equation}
\partial_{t}\vec{M}+\partial_{i}\vec{Q}_{i}=-\gamma(\vec{M}\times\vec{H})+\vec{N}
\label{dvecM/dt-2b*}
\end{equation}
on setting 
\begin{equation}
\vec{Q}_{i}=\gamma(\hbar/2)\hat{M}(j_{\downarrow i}-j_{\uparrow i}),
\label{Q-2b}
\end{equation}
where $\vec{Q}_{i}$ is the magnetization flux density ($i$ is the real space index), and 
\begin{equation}
\vec{N}=\vec{\Omega}\times\vec{M}+\gamma(\hbar/2)(j_{\downarrow i}-j_{\uparrow i})\partial_{i}\hat{M}-2S\gamma(\hbar/2)\hat{M},
\label{N-2b}
\end{equation}
where $\vec{N}$ (a source) is the volume rate of change of magnetization due to torques associated with a lack of thermal equilibrium.

From (\ref{Q-2b}) and (\ref{N-2b}) the net non-Larmor spin transfer torque of (\ref{N'}) is given by
\begin{equation}
\vec{N}'=\vec{\Omega}\times\vec{M}-\gamma(\hbar/2)\hat{M}\partial_{i}(j_{\downarrow i}-j_{\uparrow i})-2S\gamma(\hbar/2)\hat{M},
\label{N'-2b}
\end{equation}
Thus the only transverse part of $\vec{N}'$ comes from the $\vec{\Omega}\times\vec{M}$ term. 

Once the difference in units are accounted for, $\vec{Q}_{i}$ above is equivalent to Eq.~(8) of Ref.~\onlinecite{SXZ04}; there, both the magnetization and the magnetization flux densities are measured in units of $\gamma$, with $|g|=2$, and $\hat{u}$ is employed for the direction of the magnetization.  Moreover, the second term of $\vec{N}$ in (\ref{N-2b}) is the same as the adiabatic spin torque density of Ref.~\onlinecite{SSDZ}.  This adiabatic spin torque is enforced by the condition that the magnetization and the spin quantization axis track with one another.  Note that Ref.\onlinecite{SXZ04} does not include the adiabatic spin torque density. 

In (\ref{N-2b}), the first and second terms give the transverse components in spin space, and the third term gives the longitudinal component in spin space.  Of course, we have yet to determine $\vec{\Omega}$, $S$, $j_{\downarrow i}$, or $j_{\uparrow i}$.  Below we show that $j_{\downarrow i}$ and $j_{\uparrow i}$ each have five terms, so that $Q_{\alpha i}$ has ten terms.  Moreover, $\vec{\Omega}$ has seven terms and $S$ has one term, so that $\vec{N}$ has eighteen terms.  

Note that spin angular momentum is not conserved; however, total angular momentum is conserved once one accounts for the crystal lattice angular momentum.  With $\vec{H}'=\vec{H}_{0}+\vec{H}_{an}+\vec{H}_{ex}$, the angular momentum associated with $\vec{H}_{0}$ is transferred to the source of $\vec{H}_{0}$ (an external magnet or an external current-carrying circuit); the angular momentum associated with $\vec{H}_{an}$ (either lattice or dipolar anisotropy) is transferred to the lattice; and the angular momentum associated with $\vec{H}_{ex}$ integrates to zero, because it involves the spin system interacting with itself.  $\vec{N}$ is associated with the lattice. 

\section{Rate of Heat Production}
Irreversible thermodynamics accomplishes its task by combining the equations of motion and the thermodynamics to obtain an expression for the non-negative quantity $R$ as a sum of products of fluxes (or sources) and thermodynamic ``forces'' (or their gradients).  

\subsection{Generic Conducting Magnet}
Eqs. (\ref{de/dt}), (\ref{ds/dt}), (\ref{dn/dt-sd}), and (\ref{dvecM/dt-sd}) placed in the time-derivative of (\ref{de-sd}) yields
\begin{eqnarray}
0\le R&=&-\partial_{i}\Bigl(j^{\epsilon}_{i}-Tj^{s}_{i}-\tilde\mu j^{n}_{i}+\vec{H}\cdot\vec{Q}_{i}\Bigr)\cr
&&-j^{s}_{i}\partial_{i}T-j^{n}_{i}\partial_{i}\tilde\mu+\vec{Q}_{i}\cdot\partial_{i}\vec{H}+\vec{N}\cdot\vec{H}.
\label{R1-sd}
\end{eqnarray}
Here $j^{s}_{i}$, $j^{n}_{i}$, and $\vec{Q}_{i}$ are thermodynamic fluxes and $\partial_{i}\vec{H}$, $\partial_{i}T$, and $\partial_{i}\tilde\mu$ are thermodynamic forces; further, $\vec{N}$ is a thermodynamic source, and $\vec{H}$ is a thermodynamic force.  In equilibrium all of the thermodynamic forces are zero, and thus there is no entropy production.

Each of the four non-divergence terms in (\ref{R1-sd}) has a clear physical interpretation as a source of heating: the first term to thermal conduction, the second to electrical conduction, the third to magnetic diffusion (or conduction), and the fourth to (local) spin damping. 

\subsection{Two-band Magnet}
Eqs. (\ref{de/dt}), (\ref{ds/dt}), and (\ref{dnup/dt})-(\ref{dhatM/dt}) placed in the time-derivative of (\ref{de-2b}) yields
\begin{eqnarray}
0\le R&=&T(\partial_{t}s+\partial_{i}j^{s}_{i})\cr
&=&-\partial_{i}\Bigl(j^{\epsilon}_{i}-Tj^{s}_{i}-\mu^{*}_{\uparrow} j_{\uparrow i}-\mu^{*}_{\downarrow} j_{\downarrow i}\Bigr)\cr
&&-j^{s}_{i}\partial_{i}T-j_{\uparrow i}\partial_{i}\mu^{*}_{\uparrow}-j_{\downarrow i}\partial_{i}\mu^{*}_{\downarrow}\cr
&&-(\mu^{*}_{\uparrow}-\mu^{*}_{\downarrow})S+\vec{\Omega}\cdot(\vec{M}\times\vec{H}).
\label{R1-2b}
\end{eqnarray}
Here $j^{s}_{i}$, $j_{\uparrow i}$, and $j_{\downarrow i}$ are thermodynamic fluxes and $\partial_{i}T$, $\partial_{i}\mu^{*}_{\uparrow}$, and $\partial_{i}\mu^{*}_{\downarrow}$  are thermodynamic forces; further, $S$ and $\vec{\Omega}$ are thermodynamic sources and $(\mu^{*}_{\uparrow}-\mu^{*}_{\downarrow})$ and $\vec{M}\times\vec{H}$ are thermodynamic forces. 

Each of the five non-divergence terms in (\ref{R1-2b}) has a clear physical interpretation as a source of heating: the first term to thermal conduction, the second and third to (spin-dependent) electrical conduction, the fourth to (local) longitudinal magnetic damping, and the fifth to (local) transverse magnetic damping. 

\section{Sources and Fluxes}
\subsection{Generic Conducting Magnet}
This model has been treated in the absence of terms associated with the conducting electrons.\cite{RivkinSaslow}

{\bf Energy Flux $j^{\epsilon}_{i}$.}
The energy flux is given by constraining the divergence to be zero (up to an arbitrary curl), which leads to 
\begin{equation}
j^{\epsilon}_{i}=Tj^{s}_{i}+\tilde\mu j^{n}_{i}+\vec{H}\cdot\vec{Q}_{i}.
\label{je-sd}
\end{equation}

We now must express each flux and source as the sum over the suitably weighted ``forces''  $\partial_{i}T$, $\partial_{i}\tilde\mu$, $\vec{H}$, and $\partial_{i}\vec{H}$, all of which are zero in equilibrium.  The coefficients may be constructed from the ``order parameters'' of the equilibrium state, $\vec{M}$ and $\partial_{i}\vec{M}$.  The vector nature of the fluxes must be respected (including their properties under both real space and spin space rotations).  

{\bf Entropy Flux $j^{s}_{i}$.}
The entropy flux, a vector in real space whose non-dissipative part is odd under time-reversal ${\cal T}$, takes the form
\begin{eqnarray}
j^{s}_{i}&=&-\frac{\kappa}{T}\partial_{i}T -L_{sn}\partial_{i}\tilde\mu -L_{sQ}\vec{M}\cdot\partial_{i}\vec{H}\cr
&&-L_{sM1}(\partial_{i}\vec{M})\cdot\vec{H}-L_{sM2}(\hat{M}\cdot\partial_{i}\vec{M})(\hat{M}\cdot\vec{H})\cr
&&-L'_{sM}\partial_{i}\vec{M}\cdot(\vec{M}\times\vec{H}).
\label{js-sd}
\end{eqnarray}
There are six terms.  The terms with unprimed coefficients are even under time-reversal ${\cal T}$, signifying dissipation.  The term with a primed coefficient is odd under ${\cal T}$ (signifying no dissipation). $\kappa$ is the usual thermal conductivity.  $L_{sn}$ has units of [diffusion-constant-density/ temperature].  $L_{sQ}$, $L_{sM1}$, and $L_{sM2}$ have units of [velocity-length/temperature], or [diffusion-constant/temperature].  $L'_{sM}$ has units of [diffusion-constant-density/ temperature-magnetization]. 

{\bf Number Flux $j^{n}_{i}$.}
The number flux, like the entropy flux a vector in real space whose non-dissipative part is odd under time-reversal ${\cal T}$, takes the form
\begin{eqnarray}
j^{n}_{i}&=&-\frac{\sigma}{e^{2}}\partial_{i}\tilde\mu -L_{ns}\partial_{i}T 
-L_{nQ} \vec{M}\cdot\partial_{i}\vec{H}\cr
&&-L_{nM1}(\partial_{i}\vec{M})\cdot\vec{H}-L_{nM2}(\hat{M}\cdot\partial_{i}\vec{M})(\hat{M}\cdot\vec{H})\cr
&&-L'_{nM}\partial_{i}\vec{M}\cdot(\vec{M}\times\vec{H}).
\label{jn-sd}
\end{eqnarray}
There are six terms.  The terms with unprimed coefficients are even under time-reversal ${\cal T}$, signifying dissipation.  The terms with a primed coefficient is odd under ${\cal T}$ (signifying no dissipation).  $\sigma$ is the usual electrical conductivity.  The last three terms we will later interpret as {\it bulk spin pumping} terms, and we will relate them to the {\it bulk spin torque} terms, to be discussed shortly.  $L_{ns}$ has units of [diffusion-constant-density/ temperature].  $L_{nQ}$, $L_{nM1}$, and $L_{nM2}$ have units of  [diffusion-constant/energy].  $L'_{nM}$ has units of [diffusion-constant-density/ energy-magnetization].

{\bf Magnetization Flux $\vec{Q}_{i}$.}
The magnetization flux, whose non-dissipative part is even under time-reversal, takes the form
\begin{eqnarray}
\vec{Q}_{i}&=&C_{\parallel}\vec{M}(\hat{M}\cdot\partial_{i}\vec{H})-C_{\perp}\vec{M}\times(\hat{M}\times\partial_{i}\vec{H})\cr
&&-C'\vec{M}\times\partial_{i}\vec{H}+L_{Qs}\vec{M}\partial_{i}T+L_{Qn}\vec{M}\partial_{i}\tilde\mu\cr
&&+L_{QN1}\partial_{i}\vec{M}(\hat{M}\cdot\vec{H})+L_{QN2}(\hat{M}\cdot\partial_{i}\vec{M})\vec{H}\cr
&&+L_{QN3}\hat{M}(\partial_{i}\vec{M}\cdot\vec{H})+L_{QN4}\hat{M}(\hat{M}\cdot\partial_{i}\vec{M})(\hat{M}\cdot\vec{H})\cr
&&+L'_{QN1}\partial_{i}\vec{M}\times\vec{H}+L'_{QN2}\hat{M}[(\hat{M}\times\partial_{i}\vec{M})\cdot\vec{H}]\cr
&&+L'_{QN3}(\hat{M}\times\partial_{i}\vec{M})(\hat{M}\cdot\vec{H})\cr
&&+L'_{QN4}(\hat{M}\cdot\partial_{i}\vec{M})(\hat{M}\times\vec{H}).
\label{Q-sd}
\end{eqnarray}
There are thirteen terms.  The terms with unprimed coefficients are odd under time-reversal ${\cal T}$, signifying dissipation.  The terms with primed coefficients are even under ${\cal T}$ (signifying no dissipation).  The terms involving $\vec{H}$ are discussed in Ref.~(\onlinecite{RivkinSaslow}), as is the term involving $\partial_{i}T$.   $C_{\parallel}$ and $C_{\perp}$ have dimensions of a diffusion constant divided by a magnetization-squared, and indeed they represent longitudinal and transverse diffusion.  $C'$ has the dimensions of [diffusion-constant/magnetization].  $L_{Qn}$ has units of [diffusion-constant/energy].  The term associated with $L_{Qn}$, when the divergence is taken, will lead to a {\it bulk spin torque} term.  All of the terms in $L_{QN}$ and $L'_{QN}$ have units of [diffusion-constant/field], with field in tesla.    Note that, for near-saturation of the magnetization, the terms $\hat{M}\cdot\partial_{i}\vec{M}$ will be small, so that only $L_{QN1}$ and $L_{QN3}$ will yield possibly important new terms in the damping. 

{\bf Spin Torque Density $\vec{N}$.}
The spin torque density, whose non-dissipative part is even under time-reversal, takes the form
\begin{eqnarray}
\vec{N}&=&A_{\parallel}\vec{M}(\vec{M}\cdot\vec{H})-A_{\perp}\vec{M}\times(\vec{M}\times\vec{H})\cr
&&+L_{Ms1}\partial_{i}\vec{M}\partial_{i}T+L_{Ms2}\hat{M}(\hat{M}\cdot\partial_{i}\vec{M})\partial_{i}T\cr
&&+L'_{Ms}(\hat{M}\times\partial_{i}\vec{M})\partial_{i}T+L'_{Mn}(\hat{M}\times\partial_{i}\vec{M})\partial_{i}\tilde\mu\cr
&&+L_{Mn1}\partial_{i}\vec{M}\partial_{i}\tilde\mu+L_{Mn2}\hat{M}(\hat{M}\cdot\partial_{i}\vec{M})\partial_{i}\tilde\mu\cr
&&+L_{NQ1}\hat{M}(\partial_{i}\vec{M}\cdot\partial_{i}\vec{H})+L_{NQ2}(\hat{M}\cdot\partial_{i}\vec{M})\partial_{i}\vec{H}\cr
&&+L_{NQ3}\partial_{i}\vec{M}(\hat{M}\cdot\partial_{i}\vec{H})+L_{NQ4}\hat{M}(\hat{M}\cdot\partial_{i}\vec{M})(\hat{M}\cdot\partial_{i}\vec{H})\cr
&&+L'_{NQ1}\partial_{i}\vec{M}\times\partial_{i}\vec{H}+L'_{NQ2}(\hat{M}\times\partial_{i}\vec{M})(\hat{M}\cdot\partial_{i}\vec{H})\cr
&&+L'_{NQ3}\hat{M}(\hat{M}\times\partial_{i}\vec{M})\cdot\partial_{i}\vec{H}\cr
&&+L'_{NQ4}(\hat{M}\cdot\partial_{i}\hat{M})\hat{M}\times\partial_{i}\vec{H}.
\label{N-sd}
\end{eqnarray}
There are sixteen terms.  The terms with unprimed coefficients are odd under time-reversal ${\cal T}$, signifying dissipation.  The terms with primed coefficients are even under ${\cal T}$ (signifying no dissipation).  The terms involving $\partial_{i}\vec{H}$ have been discussed in Ref.~(\onlinecite{RivkinSaslow}), as has the term involving $\partial_{i}T$.  The terms in $L_{Mn1},L_{Mn2},L'_{Mn}$ we interpret as {\it bulk spin torque} terms.  $A_{\parallel}$ and $A_{\perp}$ have units of [1/time-(magnetization)$^{2}$] for $\vec{H}$ given in A/m; or they must include a factor of $1/\mu_{0}$ if $\vec{H}$ is given in T.  They represent longitudinal and transverse damping; in terms of the Landau-Lifshitz parameter $\lambda$, one has $A_{\perp}=\lambda/M$.  $L_{Ms1}$, and $L_{Ms2}$ have units of [velocity-length/temperature], or [diffusion-constant/temperature].  $L'_{Ms}$ has units of [diffusion-constant-density/ temperature-magnetization].  $L_{Mn1}$, and $L_{Mn2}$ have units of  [diffusion-constant/energy].  $L'_{Mn}$ has units of [diffusion-constant-density/ energy-magnetization].  All of the terms in $L_{NQ}$ and $L'_{NQ}$ have units of [diffusion-constant/field], with field in tesla.  

{\bf Rate of Entropy Production $R$.}
The rate of entropy production is strictly even under time-reversal.  A total of forty-one terms can contribute to $R$ when the above equations are substituted to find $R$.  The term involving $C'$ is identically zero.  There are six diagonal terms once the longitudinal and transverse parts of $\vec{Q}_{i}$ and $\vec{N}$ are accounted for.  The remaining thirty-four terms are cross-terms that occur in pairs.  We find that
\begin{eqnarray}
R&=&\frac{\kappa}{T}(\partial_{i}T)^{2}+\frac{\sigma}{e^{2}}(\partial_{i}\tilde\mu)^{2}
+(L_{sn}+L_{ns})\partial_{i}T\partial_{i}\tilde\mu\cr
&&+C_{\parallel}(\vec{M}\cdot\partial_{i}\vec{H})^{2}+C_{\perp}(\vec{M}\times\partial_{i}\vec{H})^{2}\cr
&&+A_{\parallel}(\vec{M}\cdot\vec{H})^{2}+A_{\perp}(\vec{M}\times\vec{H})^{2}\cr
&&+(L_{sQ}+L_{Qs})\vec{M}\cdot\partial_{i}\vec{H}\partial_{i}T\cr
&&+(L_{nQ}+L_{Qn})\vec{M}\cdot\partial_{i}\vec{H}\partial_{i}\tilde\mu\cr
&&+(L_{sM1}+L_{Ms1})\partial_{i}\vec{M}\cdot\vec{H}\partial_{i}T\cr
&&+(L_{nM1}+L_{Mn1})\partial_{i}\vec{M}\cdot\vec{H}\partial_{i}\tilde\mu\cr
&&+(L_{sM2}+L_{Ms2})(\hat{M}\cdot\partial_{i}\vec{M})(\hat{M}\cdot\vec{H})\partial_{i}T\cr
&&+(L_{nM2}+L_{Mn2})(\hat{M}\cdot\partial_{i}\vec{M})(\hat{M}\cdot\vec{H})\partial_{i}\tilde\mu\cr
&&+(L'_{sM}+L'_{Ms})(\hat{M}\times\partial_{i}\vec{M})\cdot\vec{H}\partial_{i}T\cr
&&+(L'_{nM}+L'_{Mn})(\hat{M}\times\partial_{i}\vec{M})\cdot\vec{H}\partial_{i}\tilde\mu\cr
&&+(L_{QN1}+L_{NQ1})(\hat{M}\cdot\vec{H})(\partial_{i}\vec{M}\cdot\partial_{i}\vec{H})\cr
&&+(L_{QN2}+L_{NQ2})(\hat{M}\cdot\partial_{i}\vec{M})(\vec{H}\cdot\partial_{i}\vec{H})\cr
&&+(L_{QN3}+L_{NQ3})(\hat{M}\cdot\partial_{i}\vec{H})(\partial_{i}\vec{M}\cdot\vec{H})\cr
&&+(L_{QN4}+L_{NQ4})(\hat{M}\cdot\partial_{i}\vec{M})(\hat{M}\cdot\vec{H})(\hat{M}\cdot\partial_{i}\vec{H})\cr
&&+(L'_{QN1}+L'_{NQ1})(\partial_{i}\vec{M}\times\vec{H}\cdot\partial_{i}\vec{H})\cr
&&+(L'_{QN2}+L'_{NQ2})(\hat{M}\times\partial_{i}\vec{M}\cdot\vec{H})(\hat{M}\cdot\partial_{i}\vec{H})\cr
&&+(L'_{QN3}+L'_{NQ3})(\hat{M}\cdot\vec{H})(\hat{M}\times\partial_{i}\vec{M}\cdot\partial_{i}\vec{H})\cr
&&+(L'_{QN4}+L'_{NQ4})(\hat{M}\cdot\partial_{i}\vec{M})(\hat{M}\times\vec{H}\cdot\partial_{i}\vec{H}).
\label{R2-sd}
\end{eqnarray}
Recall that $\vec{H}\ne\vec{H}'$, and that $\vec{H}=\vec{0}$ in equilibrium, so that there are no dissipative terms associated with $\vec{H}$ when the magnetization is in equilibrium. 

To ensure that $R$ is invariant under time-reversal the non-dissipative cross-terms (which are odd under ${\cal T}$) must be eliminated.  This is done by imposing the conditions 
\begin{eqnarray}
L'_{sM}=-L'_{Ms}, \qquad L'_{nM}=-L'_{Mn},\cr
L'_{NQi}=-L'_{QNi}, i=1,4
\end{eqnarray}

In addition, in order to satisfy the Onsager relations for dissipative cross-terms (which are even under ${\cal T}$) we must impose the conditions 
\begin{eqnarray}
L_{ns}=L_{ns}, \quad L_{sQ}=L_{Qs}, \quad L_{nQ}=L_{Qn},\cr
L_{sM1}=L_{Ms1}, \quad L_{sM2}=L_{Ms2},\cr
L_{nM1}=L_{Mn1}, \quad L_{nM2}=L_{Mn2},\cr 
L_{QNi}=L_{NQi}, i=1,4.
\end{eqnarray}
We then have
\begin{eqnarray}
R&=&\frac{\kappa}{T}(\partial_{i}T)^{2}+\frac{\sigma}{e^{2}}(\partial_{i}\tilde\mu)^{2}
+2L_{sn}\partial_{i}T\,\partial_{i}\tilde\mu\cr
&&+C_{\parallel}(\vec{M}\cdot\partial_{i}\vec{H})^{2}+C_{\perp}(\vec{M}\times\partial_{i}\vec{H})^{2}\cr
&&+A_{\parallel}(\vec{M}\cdot\vec{H})^{2}+A_{\perp}(\vec{M}\times\vec{H})^{2}\cr
&&+2L_{sQ}\vec{M}\cdot\partial_{i}\vec{H}\partial_{i}T
+2L_{nQ}\vec{M}\cdot\partial_{i}\vec{H}\partial_{i}\tilde\mu\cr
&&+2L_{sM1}\partial_{i}\vec{M}\cdot\vec{H}\partial_{i}T
+2L_{nM1}\partial_{i}\vec{M}\cdot\vec{H}\partial_{i}\tilde\mu\cr
&&+2L_{sM2}(\hat{M}\cdot\partial_{i}\vec{M})(\hat{M}\cdot\vec{H})\partial_{i}T\cr
&&+2L_{nM2}(\hat{M}\cdot\partial_{i}\vec{M})(\hat{M}\cdot\vec{H})\partial_{i}\tilde\mu\cr
&&+2L_{QN1}(\hat{M}\cdot\vec{H})(\partial_{i}\vec{M}\cdot\partial_{i}\vec{H})\cr
&&+2L_{QN2}(\hat{M}\cdot\partial_{i}\vec{M})(\vec{H}\cdot\partial_{i}\vec{H})\cr
&&+2L_{QN3}(\hat{M}\cdot\partial_{i}\vec{H})(\partial_{i}\vec{M}\cdot\vec{H})\cr
&&+2L_{QN4}(\hat{M}\cdot\partial_{i}\vec{M})(\hat{M}\cdot\vec{H})(\hat{M}\cdot\partial_{i}\vec{H})
\label{R3-sd}
\end{eqnarray}

{\bf Dissipative Nature of Adiabatic Spin Torque.}  All the terms in (\ref{R3-sd}), being even under time-reversal, are dissipative.  We first call attention to the third term, due to the thermoelectric effect, involving the product of thermodynamic forces $\partial_{i}T\,\partial_{i}\tilde\mu$.  Had $\partial_{i}\tilde\mu$  been replaced (up to a proportionality constant) by $j^{n}_{i}$, to give the form $\partial_{i}T\,j^{n}_{i}$, which is a product of a thermodynamic force and a flux, one would have a term that is apparently odd under time-reversal.  One therefore might conclude that such a term is non-dissipative.  Such reasoning would be incorrect.  In determining the time-reversal properties of terms that might contribute to the rate of heating, one must work either with products of forces or fluxes; we have done the former.  

Similarly, in (\ref{R3-sd}) the terms in $L_{Mn1}$ and $L_{Mn2}$, which are due both to adiabatic {\it bulk spin torque} and adiabatic {\it bulk spin pumping} terms, produce dissipation.  Likewise the terms in $L_{nQ}$, which also produce a (longitudinal) bulk spin pumping term and a longitudinal spin flux, produce dissipation.  

\subsection{Two-band Magnet}

{\bf Energy Flux $j^{\epsilon}_{i}$.}
The energy flux is given by constraining the divergence to be zero (up to an arbitrary curl), which leads to 
\begin{equation}
j^{\epsilon}_{i}=Tj^{s}_{i}-\mu^{*}_{\uparrow} j_{\uparrow i}-\mu^{*}_{\downarrow} j_{\downarrow i}.
\label{je-2b}
\end{equation}

We now must express each flux and source as the sum over the suitably weighted ``forces''  $\tilde\mu^{*}_{\uparrow}-\tilde\mu^{*}_{\downarrow}$, $\partial_{i}T$, $\partial_{i}\tilde\mu^{*}_{\uparrow}$, $\partial_{i}\tilde\mu^{*}_{\downarrow}$, and $\vec{M}\times\vec{H}$, all of which are zero in equilibrium.  The coefficients may be constructed from the ``order parameters'' of the equilibrium state, $\vec{M}$ and $\partial_{i}\vec{M}$.  The vector nature of the fluxes must be respected (including their properties under both real space and spin space rotations).  

{\bf Decay Rate $S$.}
The intrinsic signatures under time-reversal ${\cal T}$ of $n_{\uparrow}$ and $n_{\downarrow}$ are even, so that the intrinsic signatures under ${\cal T}$ of their time-derivatives are odd.  The decay rate $S$ is associated with both $\partial_{t}n{\uparrow}$ and $\partial_{t}n{\downarrow}$; hence the non-dissipative part of $S$ is odd under ${\cal T}$.

For $S$, the only possible form is 
\begin{equation}
S=-\alpha(\mu^{*}_{\uparrow}-\mu^{*}_{\downarrow}),
\label{S}
\end{equation}
with $\alpha$ a material-dependent constant havings units of (J-m$^{3}$-s)$^{-1}$.  No other form is allowed because $S$ is a scalar in both real space and spin space.  One might think that the ``order parameter'' $\vec{M}$ could be multiplied by the thermodynamic ``force'' $\vec{M}\times\vec{H}$ to obtain a scalar, but that dot product is identically zero.  Eq.~(\ref{S}) is even under time-reversal, and therefore is dissipative. 

For small deviations from equilibrium, we have 
\begin{eqnarray}
S\approx-\alpha\biggr(\frac{\partial\mu^{*}_{\uparrow}}{\partial n_{\uparrow}}{\delta n_{\uparrow}}
-\frac{\partial\mu^{*}_{\downarrow}}{\partial n_{\downarrow}}{\delta n_{\downarrow}}\biggl)
=-\frac{\delta n_\uparrow}{\tau_{\uparrow}}+\frac{\delta n_\downarrow}{\tau_{\downarrow}},\\
\frac{1}{\tau_{\uparrow}}\equiv\alpha\frac{\partial\mu^{*}_{\uparrow}}{\partial n_{\uparrow}}, \quad
\frac{1}{\tau_{\downarrow}}\equiv\alpha\frac{\partial\mu^{*}_{\downarrow}}{\partial n_{\downarrow}}, 
\end{eqnarray}
a result that could have been expected on physical grounds.  Note that if $\tau_{\uparrow}=\tau_{\downarrow}$, then the longitudinal part of the magnetization would have a decay rate proportional to the deviation in the longitudinal magnetization itself.  However, since under most circumstances local electroneutrality enforces $\delta n_{\uparrow}=-\delta n_{\downarrow}$, so $\delta M\sim\delta n_{\uparrow}$, this normally will be the case anyway.  In this case we can think of $S$ as determining $T_{1}$ processes. 

{\bf Entropy Flux $j^{s}_{i}$.}
The entropy flux, a vector in real space whose reversible part is odd under time-reversal ${\cal T}$, takes the form
\begin{eqnarray}
j^{s}_{i}&=&-\frac{\kappa}{T}\partial_{i}T -L_{s\uparrow}\partial_{i}\mu^{*}_{\uparrow}-L_{s\downarrow}\partial_{i}\mu^{*}_{\downarrow}\cr
&&-L_{sM}(\vec{M}\times\partial_{i}\vec{M})\cdot(\vec{M}\times\vec{H})\cr
&&-L'_{sM}\partial_{i}\vec{M}\cdot(\vec{M}\times\vec{H}).
\label{js1}
\end{eqnarray}
There are five terms.  The terms with unprimed coefficients are even under time-reversal, signifying dissipation.  The term with a primed coefficient is odd under time-reversal (signifying no dissipation).  
$\kappa$ is the usual thermal conductivity.  $L_{s\uparrow}$ and $L_{s\downarrow}$, associated with the second and third terms, give a well-known electrothermal effect, whereby a chemical potential gradient can drive an entropy current; they have units of a diffusion constant divided by [temperature].  The last two terms, which are new, have the same symmetry as corresponding terms in the number flux, which we now discuss.  They imply that spin dynamics can drive an entropy current.  We may call this spin-pumping of an entropy current.  $L_{s\vec{M}}$ has units of a diffusion constant divided by [magnetization$^{2}$-temperature], and $L'_{s\vec{M}}$ has units of a diffusion constant divided by [magnetization-temperature].
Note that when the fifth term is multiplied by $\partial_{i}T$ in (\ref{R1-2b}) for $R$, unlike $R$ the product is odd under ${\cal T}$; it thus must be canceled by another term (to be discussed below) or it must be zero.  

{\bf Number Fluxes $j_{\uparrow i}$ and $j_{\downarrow i}$.}
Each number flux, like the entropy flux, is a vector in real space whose reactive, or reversible, part is odd under time-reversal ${\cal T}$. They take the forms [cf.(\ref{js1})]:
\begin{eqnarray}
j_{\uparrow i}&=&-\frac{\sigma_{\uparrow}}{e^{2}}\partial_{i}\mu^{*}_{\uparrow}-L_{\uparrow s}\partial_{i}T 
-L_{\uparrow\downarrow}\partial_{i}\mu^{*}_{\downarrow}\cr
&&-L_{\uparrow M} (\vec{M}\times\partial_{i}\vec{M})\cdot(\vec{M}\times\vec{H})\cr
&&-L'_{\uparrow M}\partial_{i}\vec{M}\cdot(\vec{M}\times\vec{H}),\label{jup}\\
j_{\downarrow i}&=&-\frac{\sigma_{\downarrow}}{e^{2}}\partial_{i}\mu^{*}_{\downarrow}-L_{\downarrow s}\partial_{i}T 
-L_{\downarrow\uparrow}\partial_{i}\mu^{*}_{\uparrow}\cr
&&-L_{\downarrow M} (\vec{M}\times\partial_{i}\vec{M})\cdot(\vec{M}\times\vec{H})\cr
&&-L'_{\downarrow M}\partial_{i}\vec{M}\cdot(\vec{M}\times\vec{H}).\label{jdown}
\end{eqnarray}
Each of these have five terms.  The terms with unprimed coefficients are even under time-reversal, signifying dissipation.  The term with a primed coefficient is odd under time-reversal (signifying no dissipation).  $\sigma_{\uparrow\downarrow}$ gives the respective electrical conductivities.  $L_{s\uparrow}$ and $L_{s\downarrow}$, associated with the second and third terms, give a well-known electrothermal effect, whereby a chemical potential gradient can drive an entropy current; they have units of a diffusion constant divided by [temperature]; the terms $L_{\downarrow\uparrow}$ and $L_{\uparrow\downarrow}$ have the same units.  The last two terms in both (\ref{jup}) and (\ref{jdown}), which are new, have the same symmetry as corresponding terms in the number flux, which we now discuss.  They imply that spin dynamics can drive a current.  We call this spin-pumping of current.  $L_{\uparrow\vec{M}}$ and $L_{\downarrow\vec{M}}$ have units of a diffusion constant divided by [magnetization$^{2}$-energy], and $L'_{\uparrow\vec{M}}$ and $L'_{\downarrow\vec{M}}$  has units of a diffusion constant divided by [magnetization-energy].   Note that when the fifth terms are respectively multiplied by $\partial_{i}n_{\uparrow}$ and $\partial_{i}n_{\downarrow}$ in (\ref{R1-2b}) for $R$, unlike $R$ the product is odd under ${\cal T}$, and thus these terms must be canceled by another cross-term (discussed below) or they must be zero.

{\bf Non-equilibrium Rotation Rate $\vec{\Omega}$.}
The reactive, or reversible, part of $\vec{\Omega}$ is odd under ${\cal T}$.  $\vec{\Omega}$ bears the brunt of the complexity of the Onsager coefficients.  In detail $\vec{\Omega}$ is given by
\begin{eqnarray}
\vec{\Omega}&=&\lambda\hat{M}\times\vec{H}+L_{Ms}(\vec{M}\times\partial_{i}\vec{M})\partial_{i}T+L'_{Ms}\partial_{i}\vec{M}\partial_{i}T\cr
&&+L_{M\uparrow}(\vec{M}\times\partial_{i}\vec{M})\partial_{i}\mu^{*}_{\uparrow}+L'_{M\uparrow}\partial_{i}\vec{M}\partial_{i}\mu^{*}_{\uparrow}\cr
&&+L_{M\downarrow}(\vec{M}\times\partial_{i}\vec{M})\partial_{i}\mu^{*}_{\downarrow}+L'_{M\downarrow}\partial_{i}\vec{M}\partial_{i}\mu^{*}_{\downarrow}.
\label{Omega}
\end{eqnarray}
There are seven terms.  The terms with unprimed coefficients are even under ${\cal T}$, and thus are dissipative.  The terms with primed coefficients are odd under ${\cal T}$, and thus are not dissipative.  
The first term gives Landau-Lifshitz damping.  (We believe that any theory based on irreversible thermodynamics that has an energy term varying as $-\vec{H}M\cdot d\hat{M}$ will give a Landau-Lifshitz damping term, since then the corresponding thermodynamic force is $\hat{M}\times\vec{H}$.) 
The three terms proportional to $\vec{M}\times\partial_{i}\vec{M}$ are even under ${\cal T}$, so they are dissipative.  They correspond to spin torque by electric current and by entropy current.  Note that there are two types of spin torque, corresponding to the two types of spins; this is a new result.  Both of these correspond to what has been called adiabatic spin torque.  The term driven by entropy current is new.
The three terms proportional to $\partial_{i}\vec{M}$ are odd under ${\cal T}$, so they are non-dissipative.  They, too, correspond to spin torque by electric current and by entropy current.  Again there are two types of spin torque, corresponding to the two types of spins; this is a new result.  Both of these correspond to what has been called non-adiabatic spin torque.  The term driven by entropy current is new.

{\bf Rate of Heat Production $R$.}
The rate of entropy production is strictly even under time-reversal.  We now rewrite (\ref{R1-2b}) in light of (\ref{je-2b}), which eliminates the divergence term, and in light of the equations for the various thermodynamic fluxes and sources.  We then have
\begin{eqnarray}
R&=&\,\alpha(\mu^{*}_{\uparrow}-\mu^{*}_{\uparrow})^{2}+\frac{\kappa}{T}(\partial_{i}T)^{2}+\frac{\sigma_{\uparrow}}{e^{2}}(\partial_{i}\mu^{*}_{\uparrow})^{2}+\frac{\sigma_{\downarrow}}{e^{2}}(\partial_{i}\mu^{*}_{\downarrow})^{2}\cr
&&+\frac{\lambda}{M}(\vec{M}\times\vec{H})^{2}+(\partial_{i}T)(\partial_{i}\mu^{*}_{\uparrow})(L_{s{\uparrow}}+L_{{\uparrow}s})\cr
&&+(\partial_{i}T)(\partial_{i}\mu^{*}_{\downarrow})(L_{s{\downarrow}}+L_{{\downarrow}s})
+(\partial_{i}\mu^{*}_{\uparrow})(\partial_{i}\mu^{*}_{\downarrow})(L_{{\uparrow}{\downarrow}}+L_{{\downarrow}{\uparrow}})\cr
&&+(\vec{M}\times\partial_{i}\vec{M})\cdot(\vec{M}\times\vec{H})	\bigl[(L_{sM}+L_{Ms})\partial_{i}T\cr
&&\phantom{m}+(L_{{\uparrow}M}+L_{M{\uparrow}})\partial_{i}\mu^{*}_{\uparrow}
+(L_{{\downarrow}M}+L_{M{\downarrow}})\partial_{i}\mu^{*}_{\downarrow}	\bigr]\cr
&&+(\partial_{i}\vec{M})\cdot(\vec{M}\times\vec{H})	\bigl[(L'_{sM}+L'_{Ms})\partial_{i}T\cr
&&\phantom{m}+(L'_{{\uparrow}M}+L'_{M{\uparrow}})\partial_{i}\mu^{*}_{\uparrow}
+(L'_{{\downarrow}M}+L'_{M{\downarrow}})\partial_{i}\mu^{*}_{\downarrow}	\bigr].
\label{R2}
\end{eqnarray}

To ensure that $R$ is invariant under time-reversal the non-dissipative cross-terms (which are odd under ${\cal T}$) must be eliminated.  This is done by imposing the conditions 
\begin{equation}
L'_{Ms}=-L'_{sM}, \quad L'_{M{\uparrow}}=-L'_{\uparrow M}, \quad L'_{M{\downarrow}}=-L'_{\downarrow M}. 
\end{equation}
In addition, in order to satisfy the Onsager relations for dissipative cross-terms (which are even under ${\cal T}$) we must impose the conditions 
\begin{eqnarray}
L_{{\uparrow}s}=L_{s{\uparrow}}, \quad L_{{\downarrow} s}=L_{s{\downarrow}}, \quad L_{{\uparrow} {\downarrow}}=L_{{\downarrow}{\uparrow}},\cr
L_{Ms}=L_{sM}, \quad L_{M{\uparrow}}=L_{\uparrow M}, \quad L_{M{\downarrow}}=L_{\downarrow M}. 
\end{eqnarray}
We then have 
\begin{eqnarray}
R&=&\alpha(\mu^{*}_{\uparrow}-\mu^{*}_{\uparrow})^{2}+\frac{\kappa}{T}(\partial_{i}T)^{2}+\frac{\sigma_{\uparrow}}{e^{2}}(\partial_{i}\mu^{*}_{\uparrow})^{2}+\frac{\sigma_{\downarrow}}{e^{2}}(\partial_{i}\mu^{*}_{\downarrow})^{2}\cr
&&+\frac{\lambda}{M}(\vec{M}\times\vec{H})^{2}+2L_{{\uparrow}{\downarrow}}(\partial_{i}\mu^{*}_{\uparrow})(\partial_{i}\mu^{*}_{\downarrow})\cr
&&+2L_{s{\uparrow}}(\partial_{i}T)(\partial_{i}\mu^{*}_{\uparrow})
+2L_{s{\downarrow}}(\partial_{i}T)(\partial_{i}\mu^{*}_{\downarrow})\cr
&&+2(\vec{M}\times\partial_{i}\vec{M})\cdot(\vec{M}\times\vec{H})
\,\bigl[	L_{sM}(\partial_{i}T)\cr
&&\phantom{mmmmmm}+L_{{\uparrow}M}(\partial_{i}\mu^{*}_{\uparrow})
+L_{{\downarrow}M}(\partial_{i}\mu^{*}_{\downarrow})	\bigr],
\label{R3}
\end{eqnarray}
The first five terms are diagonal in the fluxes and source.  They are always positive.  The other six terms, involving off-diagonal products of the thermodynamic forces, can have either sign, dependent upon the relative directions of these gradients.  The constraint of positive definiteness for $R$ then imposes limits the magnitudes of the off-diagonal coefficients, which we do not enumerate. 

\subsection{Derived Quantities}
We now turn to the derived quantities $j^{n}_{i}$, $Q_{\alpha i}$, and $N_{\alpha}$.

{\bf Number current density $j^{n}_{i}$.}
By (\ref{dn/dt-2b}), (\ref{jup}), and (\ref{jdown}), we have
\begin{eqnarray}
j^{n}_{i}=&&-\partial_{i}T(L_{\uparrow s}+L_{\downarrow s})
-\partial_{i}\mu^{*}_{\uparrow}(\dfrac{\sigma_{\uparrow}}{e^{2}}+L_{\downarrow\uparrow})
-\partial_{i}\mu^{*}_{\downarrow}(\dfrac{\sigma_{\downarrow}}{e^{2}}+L_{\uparrow\downarrow})\nonumber\\
&&-(\vec{M}\times\partial_{i}\vec{M})\cdot(\vec{M}\times\vec{H})(L_{\downarrow M}+L_{\uparrow M})\cr
&&-\partial_{i}\vec{M}\cdot(\vec{M}\times\vec{H})(L'_{\downarrow M}+L'_{\uparrow M}).
\label{jn-2b}
\end{eqnarray}

{\bf Magnetization Flux $\vec{Q}_{i}$.}
This quantity is the sum of ten terms, which we obtain from (\ref{Q-2b}), (\ref{jup}), and (\ref{jdown}).  We have
\begin{eqnarray}
\vec{Q}_{i}&=\gamma(\hbar/2)\hat{M}\biggl(
&\partial_{i}T(L_{\uparrow s}-L_{\downarrow s})\cr
&&+\partial_{i}\mu^{*}_{\uparrow}(\dfrac{\sigma_{\uparrow}}{e^{2}}-L_{\downarrow\uparrow})
-\partial_{i}\mu^{*}_{\downarrow}(\dfrac{\sigma_{\downarrow}}{e^{2}}-L_{\uparrow\downarrow})\nonumber\\
&&-(\vec{M}\times\partial_{i}\vec{M})\cdot(\vec{M}\times\vec{H})(L_{\downarrow M}-L_{\uparrow M})\cr
&&-\partial_{i}\vec{M}\cdot(\vec{M}\times\vec{H})(L'_{\downarrow M}-L'_{\uparrow M})	\biggr).
\label{Q2-2b}
\end{eqnarray}
The teems involving gradients of the electrochemical potentials are diffusive in nature. 

{\bf Magnetization Source $\vec{N}$.}
This quantity is the sum of eighteen terms, which we obtain from (\ref{N-2b}), (\ref{jup}), (\ref{jdown}), (\ref{S}), and (\ref{Omega}).  We have
\begin{eqnarray}
\vec{N}=&
\gamma(\hbar/2)&\partial_{i}\hat{M}\biggl(	
\partial_{i}T(L_{\uparrow s}-L_{\downarrow s})\cr
&&+\partial_{i}\mu^{*}_{\uparrow}(\frac{\sigma_{\uparrow}}{e^{2}}-L_{\downarrow\uparrow})
-\partial_{i}\mu^{*}_{\downarrow}(\frac{\sigma_{\downarrow}}{e^{2}}-L_{\uparrow\downarrow})\cr
&&-(\vec{M}\times\partial_{i}\vec{M})\cdot(\vec{M}\times\vec{H})(L_{\downarrow M}-L_{\uparrow M})\cr
&&-\partial_{i}\vec{M}\cdot(\vec{M}\times\vec{H})(L'_{\downarrow M}-L'_{\uparrow M})	\biggr)\cr
&-(\gamma\hbar)&\hat{M}(-1)\Gamma(\mu^{*}_{\uparrow}-\mu^{*}_{\downarrow})\nonumber\\
 &-\vec{M}\times&\biggl(	
\lambda\hat{M}\times\vec{H}\cr
&&+L_{Ms}(\vec{M}\times\partial_{i}\vec{M})\partial_{i}T+L'_{Ms}\partial_{i}\vec{M}\partial_{i}T\cr
&&+L_{M\uparrow}(\vec{M}\times\partial_{i}\vec{M})\partial_{i}\mu^{*}_{\uparrow}+L'_{M\uparrow}\partial_{i}\vec{M}\partial_{i}\mu^{*}_{\uparrow}\cr
&&+L_{M\downarrow}(\vec{M}\times\partial_{i}\vec{M})\partial_{i}\mu^{*}_{\downarrow}+L'_{M\downarrow}\partial_{i}\vec{M}\partial_{i}\mu^{*}_{\downarrow}\biggr).\cr
&&
\label{N2-2b}
\end{eqnarray}

To end this section we note that, by (\ref{N'-2b}), in the absence of temperature and chemical potential gradients, the non-Larmor spin torque has a transverse part given only by the Landau-Lifshitz term.  This is in contrast to the case of the generic conducting magnet, where two new coefficients can contribute to the damping, in a manner that depends on the magnetic texture.  For the two-band magnet there is no such dependence. 

\section{Current-Induced Spin Torque and Spin-Pumping of Current}
We are now prepared to discuss both the current-induced spin torque and the spin-pumping of current. 
For the two models studied we will restrict ourselves to the appropriate terms in the net current density and the spin torque.  For simplicity we will consider only the subset of terms that are relevant to spin torque and spin pumping, and for that reason we use ($\approx$) to indicate that the appropriate quantities contain certain terms, but are not restricted to those terms. 

\subsection{Spin Torque and Spin-Pumping Results}
\subsubsection{Generic Conducting Magnet}
Eq.~(\ref{jn-sd}) contains the spin-pumping of number current terms
\begin{eqnarray}
j^{n}_{i}&\approx&-L_{nM1}(\partial_{i}\vec{M})\cdot\vec{H}-L_{nM2}(\hat{M}\cdot\partial_{i}\vec{M})(\hat{M}\cdot\vec{H})\cr
&&-L'_{nM}\partial_{i}\vec{M}\cdot(\vec{M}\times\vec{H}.
\label{jn-sd2})
\end{eqnarray}
The second of these is small if the magnetization is nearly saturated, as we will assume.  

Eq.~(\ref{N-sd}) contains the current-driven spin torque terms
\begin{eqnarray}
\vec{N}&\approx&L'_{Mn}(\hat{M}\times\partial_{i}\vec{M})\partial_{i}\tilde\mu\cr
&&+L_{Mn1}\partial_{i}\vec{M}\partial_{i}\tilde\mu+L_{Mn2}\hat{M}(\hat{M}\cdot\partial_{i}\vec{M})\partial_{i}\tilde\mu.
\end{eqnarray}

Eq.~(\ref{Q-sd}) contains the current-driven spin torque terms
\begin{equation}
\vec{Q}_{i}\approx L_{Qn}\vec{M}\partial_{i}\tilde\mu.
\end{equation}

The total non-Larmor spin torque $\vec{N}'$ thus contains the current-driven terms
\begin{eqnarray}
\vec{N}'&\approx& L'_{Mn}(\hat{M}\times\partial_{i}\vec{M})\partial_{i}\tilde\mu\cr
&&+(L_{Mn1}-L_{Qn})\partial_{i}\vec{M}\partial_{i}\tilde\mu+L_{Mn2}\hat{M}(\hat{M}\cdot\partial_{i}\vec{M})\partial_{i}\tilde\mu.\cr
&&
\label{N'-sd2}
\end{eqnarray}

For the generic conducting magnet, eqs.(\ref{jn-sd2}) and (\ref{N'-sd2}) provide the basis of our later discussion of the relationship between spin torque and spin pumping. 

\subsubsection{Two-band Magnet}
In the two-band magnet, each spin-component has, in principle, its own electrochemical potential.  In some cases such a distinction can be made experimentally.  (For example, by a suitable combination of electric field and of magnetic field gradient it may be possible to produce a net spin current but zero net electric current.)  For simplicity, however, let us consider that $\mu^{*}_{\uparrow}=\mu^{*}_{\downarrow}$.  Then, by (\ref{jn-2b}), $j^{n}_{i}$ contains the field-driven terms
\begin{eqnarray}
j^{n}_{i}&\approx&
-(\vec{M}\times\partial_{i}\vec{M})\cdot(\vec{M}\times\vec{H})(L_{\downarrow M}+L_{\uparrow M})\cr
&&-\partial_{i}\vec{M}\cdot(\vec{M}\times\vec{H})(L'_{\downarrow M}+L'_{\uparrow M}).
\label{jn-2b2}
\end{eqnarray}

For the non-Larmor spin torque in the two-band magnet, by (\ref{N'-2b}) we need only the transverse terms, due to $\vec{\Omega}\times\vec{M}$.  If we set $\mu^{*}_{\uparrow}=\mu^{*}_{\downarrow}=\tilde\mu$, then (\ref{Omega}) yields the current-driven terms
\begin{eqnarray}
\vec{\Omega}&\approx&
(L_{M\uparrow}+L_{M\downarrow})(\vec{M}\times\partial_{i}\vec{M})\partial_{i}\tilde\mu\cr
&&+(L'_{M\uparrow}+L'_{M\downarrow})\partial_{i}\vec{M}\partial_{i}\tilde\mu,
\label{Omega'}
\end{eqnarray}
so that, by (\ref{N'-2b}) , 
\begin{eqnarray}
\vec{N}'&\approx&
(L_{M\uparrow}+L_{M\downarrow})M^{2}\partial_{i}\vec{M}\partial_{i}\tilde\mu\cr
&&-(L'_{M\uparrow}+L'_{M\downarrow})M^{2}(\hat{M}\times\partial_{i}\vec{M})\partial_{i}\tilde\mu,
\label{N'-2b2}
\end{eqnarray}

For the two-band magnet, eqs.(\ref{jn-2b2}) and (\ref{N'-2b2}) provide the basis of our later discussion of the relationship between spin torque and spin pumping. 

\subsubsection{Comparison with Previous Work}
Eq.~ (7) of Ref.\onlinecite{SXZ04}, using a phenomenology based on both up and down bands, has the form
\begin{equation}
j^{n*}_{i}=\frac{\sigma_{\uparrow}+\sigma_{\downarrow}}{e}{E}_{i}
-\frac{D_{\uparrow}+D_{\downarrow}}{2}\partial_{i}n
-\frac{D_{\uparrow}-D_{\downarrow}}{\hbar}\hat{M}\cdot\partial_{i}\vec{M}. 
\label{jn*}
\end{equation}

Eq.~(9) of Ref.\onlinecite{SXZ04} has the form
\begin{equation}
Q^{*}_{i}=\frac{\hbar}{2}\frac{\sigma_{\uparrow}-\sigma_{\downarrow}}{e}\hat{M}{E}_{i}
-\frac{\hbar}{2}\frac{D_{\uparrow}-D_{\downarrow}}{2}\hat{M}\partial_{i}n
-\frac{D_{\uparrow}+D_{\downarrow}}{2}\partial_{i}\vec{M}. 
\label{Q*}
\end{equation}
These forms are very similar to what we have derived, in that there are three thermodynamic forces in play.  For Ref.\onlinecite{SXZ04} they are the gradient of the voltage, the gradient of the density, and the longitudinal gradient of the magnetization.  In the two-band magnet they are the gradient of the electrochemical potentials, which depend on the voltage and on the densities of the up and down spins. 

We now turn to the net torque.  The sum of (2) and (4) of Ref.\onlinecite{SSDZ} has the form
\begin{equation}
\vec{N}^{*}=-\lambda[\hat{M}\times(\vec{M}\times\vec{H})]
-v[\partial_{i}\vec{M}-\beta\hat{M}\times\partial_{i}\vec{M}],
\label{N*}
\end{equation}
where $\beta$ is dimensionless and we expect that $\beta\ll 1$.\cite{SSDZ} 

Microscopic theories for the adiabatic spin torque give, with $P$ the polarization of the current (e.g., 0.6 for 60\% in the up band)\cite{SSDZ}
\begin{equation}
v=-\frac{Pj\mu_{B}}{eM} \qquad{\rm (model)}. 
\label{v-model}
\end{equation}

To make a proper comparison with this form, we must replace the current density $j_{i}$ (a flux) by the form it takes when driven by the ``force'' $\partial_{i}\tilde\mu$.  We also employ a less model-dependent form by introducing the constant $\xi$, with units of a diffusion constant divided by energy, and let ${j}_{i}\rightarrow(\sigma/e)\partial_{i}\tilde\mu$.  Thus we write
\begin{equation}
v=-\xi\partial_{i}\tilde\mu, 
\label{v}
\end{equation}
so (\ref{N*}) becomes
\begin{equation}
\vec{N}^{*}=-\lambda[\hat{M}\times(\vec{M}\times\vec{H})]
+\xi\partial_{i}\tilde\mu[\partial_{i}\vec{M}-\beta\hat{M}\times\partial_{i}\vec{M}].
\label{N*2}
\end{equation}

In this form, which is appropriate to irreversible thermodynamics, the adiabatic spin torque (proportional to $\partial_{i}\vec{M}$) is odd under time-reversal, opposite the even signature of a non-dissipative spin torque.   Therefore the adiabatic spin torque is dissipative, as can be seen from its contribution to the rate of heating $R$ above.  On the other hand, the non-adiabatic spin torque (proportional to $\hat{M}\times\partial_{i}\vec{M}$) is even under time-reversal, signifying that it is non-dissipative, as can be seen by its absence from $R$.  

For the above model we then have 
\begin{equation}
\xi=\frac{P\sigma\mu_{B}}{e^{2}M} \qquad{\rm (model)}.
\label{xi-model}
\end{equation}

To close this subsection we note that Ref.~\onlinecite{TBBH-RMP} uses very different methods to show that, at surfaces, spin transfer torque and spin pumping are related.  We also note that Ref.~\onlinecite{BarnesMaekawa07} uses a Berry-phase and a spin-Berry-phase to predict that, for ferromagnetic conductors, there is an effective spin-dependent emf that drives an ordinary electric current and a spin emf that drives a spin current.  

\subsubsection{Comparison with the two models}
\noindent{\bf Generic Conducting Magnet.} Comparison of (\ref{N*2}) and the generic conducting magnet result (\ref{N'-sd2}) shows that the two versions of the spin torque have the same form if 
\begin{equation}
(L_{Mn1}-L_{Qn})=\xi, \qquad L'_{Mn}=-\beta\xi.
\label{notate1}
\end{equation}

\noindent{\bf Two-band magnet.} Comparison of (\ref{N*2}) and the two-band magnet result (\ref{N'-2b2}) shows that the two versions of the spin torque have the same form if 
\begin{equation}
(L_{M\uparrow}+L_{M\downarrow})M^{2}=\xi, \quad (L'_{M\uparrow}+L'_{M\downarrow})M^{2}=\beta\xi.
\label{notate2}
\end{equation}

Note that the generic (``g'') conducting magnet and the two-band (``2b'') magnet have transport coefficients with different dimensionality.  By (\ref{notate1}) and (\ref{notate2}), for later purposes we write 
\begin{equation}
L_{g}\approx L_{2b}M^{2}\approx \xi.
\end{equation}  

\subsection{Estimates}

For purposes of estimation we will employ Table~1 of Ref.~\onlinecite{SXZ04}, which for Co gives $P=0.6$, $M=1.45\times10^{6}$A/m and implies that $\sigma=\sigma_{\uparrow}+\sigma_{\downarrow}=3.4\times10^{7}$/$\Omega$-m.  With standard values of $e=1.6\times10^{-19}$~C and $\mu_{B}=9.3\times10^{-24}$A-m$^{2}$, we estimate that, for Co, $\xi\approx5.1\times10^{15}$~m$^2$/J-s.   (As indicated above, this has units of a diffusion constant over an energy.  Indeed, use of $\sigma=ne^{2}\tau/m$ and $M=Pn\mu_{B}$ gives $\xi=\tau/m$.  For a bare electron mass $m$ this corresponds to the somewhat short time of $\tau\approx4.5\times10^{-15}$~s.)  Also note that $\gamma=1.9\times10^{11}$/T-s.

Let $\delta$ be a characteristic domain wall dimension, which for purpose of estimation (in Co) we will take to be 10~nm.  Within the domain walls we may take $|\partial_{i}\vec{M}|\approx 2M/\delta$ (2 because the magnetization reverses). In order to make estimates, we will neglect the vector nature of various quantities, and not consider signs.  We will consider only the adiabatic spin torque, which we believe dominates experimentally.\cite{SSDZ}

On neglecting spin-pumping terms a field $E$ yields a current density $j$ proportional to the conductivity $\sigma$ via 
\begin{equation}
j\approx\sigma E.
\label{j-est}
\end{equation}

For non-uniform magnets there is also a current-induced spin torque $N_{ST}$ that is proportional to $j$.  We define the equivalent spin torque field $H_{ST}$ via
\begin{equation}
H_{ST}=\frac{N_{ST}}{\gamma M}.
\label{H_ST1}
\end{equation}
From (\ref{N*2}), with $L_{g}$ the appropriate Onsager coefficient (by (\ref{notate1}), for estimation purposes $L_{g}$ is like $\xi$, with units of a diffusion constant divided by an energy), we have 
\begin{equation}
N_{ST}=L_{g}\frac{eM}{\delta}E.
\label{N_ST1}
\end{equation}
Now define
\begin{equation}
R_{ST}=\frac{H_{ST}}{E},
\label{R_ST1}
\end{equation}
which is in T-m/V.  Then, by (\ref{j-est}) and (\ref{H_ST1}), eq.~(\ref{R_ST1}) yields
\begin{equation}
R_{ST}=\frac{N_{ST}}{\gamma ME}=L_{g}\frac{e}{\gamma\delta}.
\label{R_ST2}
\end{equation}

From the ``velocity'' of (\ref{v-model}) and (\ref{j-est}), we can write the spin torque terms of (\ref{N*2}) as 
\begin{equation}
N_{ST}=P\sigma E\mu_{B}\frac{1}{eM}\frac{M}{\delta}=\frac{P\sigma\mu_{B}}{e\delta}E.
\label{N_ST2}
\end{equation}

Comparison of (\ref{N_ST1}) and (\ref{N_ST2}) gives the result 
\begin{equation}
L_{g}=\frac{P\sigma\mu_{B}}{e^{2}M}.
\label{L1}
\end{equation}
This permits the estimate that, for Co, $L_{g}=5.1\times10^{15}$(m$^2$/J-s).  Then, again for Co, application of (\ref{R_ST2}) gives $R_{ST}=4\times10^{-7}$(T-m/V).  Therefore within a domain wall a true $E$-field $E_{0}=1.0\times10^{4}$V/m can cause the same torque as direct application of a magnetic field $H_{ST}=4\times10^{-3}$T.  This merely restates what is already known. 

From (\ref{jn-sd2}), an applied field $H_{0}$ that produces a torque
\begin{equation}
N=\gamma MH_{0}
\label{N-est}
\end{equation}
also produces a spin-pumping driven current within a domain wall.  With $|\partial_{i}\vec{M}|\approx2M/\delta$, this current is given by
\begin{equation}
j_{SP}=eL_{g}\frac{2M}{\delta}H_{0}.
\label{j_SP}
\end{equation}
We define the equivalent spin-pumping field $E_{SP}$ via 
\begin{equation}
E_{SP}=\frac{j_{SP}}{\sigma}.
\label{E_SP1}
\end{equation}
With the definition
\begin{equation}
R_{SP}=\frac{E_{SP}}{H_{0}}
\label{R_SP1}
\end{equation}
we find that 
\begin{equation}
R_{SP}=\frac{j_{SP}}{\sigma H_{0}}=\frac{2eM}{\delta\sigma}L_{g}.
\label{R_SP2}
\end{equation}
Note that $R_{SP}$ and $R_{ST}$ have units that are inverse to one another.

For Co this gives $R_{SP}=7.0\times10^{3}$(V/T-m).  Thus, within a domain wall a true $H$-field $H_{0}$=0.1T can cause the same current as direct application of an electric field $E_{SP}=700$(V/m).  This is a new prediction.  The corresponding voltage difference across the domain wall is on the order of $E_{SP}\delta$, or $7.0\times10^{-6}$~V.  

Consider the situation depicted in Figure~1, where $\vec{H}$ points rightward and $\partial_{x}\vec{M}$ points leftward.  With $L_{nM1}\approx\xi>0$, by (\ref{jn-sd2}) the number current density points rightward, so that (within the domain wall) the electric current density points leftward.  

\subsection{Comparison with Barnes and Maekawa}
For comparison with Ref.~\onlinecite{BarnesMaekawa07}, we define a spin-pumping emf
\begin{equation}
{\cal E}_{SP}=E_{SP}\delta.
\label{V_SP}
\end{equation}
Then the emf divided by the field, on using (\ref{R_SP1}), (\ref{R_SP2}), and (\ref{L1}), is given  by
\begin{equation}
\frac{{\cal E}_{SP}}{H_{0}}=\frac{E_{SP}}{H_{0}}\delta=\frac{2P\mu_{B}}{e}.
\label{V_SP/H}
\end{equation}
Ref.~\onlinecite{BarnesMaekawa07} finds a quantity ${\cal E}_s$ to be given by $2\mu_{B}H_{0}/e$, and that $P{\cal E}_{s}$ drives a current density.  On division by $H$, this is precisely (\ref{V_SP/H}).  We believe that this exact agreement, but not the parameter dependence, is accidental.  

\section{Summary and Discussion}
Using irreversible thermodynamics we have shown, for both a two-band magnet and a generic  conducting magnet, that current-induced spin transfer torque within a magnetic domain leads, by Onsager relations, to spin pumping of current within that domain.  For a given amount of adiabatic and non-adiabatic spin torque, the two models yield similar but distinct results for the bulk spin pumping, thus distinguishing the two models.  

This has experimental implications both for samples with conducting leads and that are electrically isolated.  For Co we estimate that within a domain wall a true $H$-field $H_{0}=0.1$~T can cause the same spin-pumped current as direct application of an electric field $E_{SP}=350$~(V/m), or, across the domain wall can cause the same effect as a voltage difference of $7.0\times10^{-6}$~V.  Correspondingly, the ratio of the effective emf to the field $H_{0}$ is, for Co, about $0.7\times10^{-4}$~V/T.  

The similarity between our results and those of Barnes and Maekawa is likely not an accident.  In the present case we have shown that the ``off-diagonal'' current-induced (``adiabatic'') spin transfer torque implies a similar ``off-diagonal'' (``adiabatic'') spin pumping of current.  Barnes and Maekawa\cite{BarnesMaekawa07} show that, in addition to being able to generate a spin transfer torque,\cite{Baz98} a spin-dependent Berry phase can generate what we have called spin pumping.  Their approach also gives a natural way to understand the associated emf-to-field ratio, $2\mu_{B}/e$.  

On the other hand, using irreversible thermodynamics, it is clear from the time-reversal properties of the thermodynamic fluxes that both the spin transfer torque and the spin pumping emf correspond to irreversible processes.  The irreversible nature of these quantities is not clear from Berry-phase approaches, where one employs currents (fluxes) as primary variables, whereas in considering experimental quantities, which correspond to thermal averaging, these terms must be considered to be driven by thermodynamic forces.   






\section{Acknowledgements}
I would like to thank M. D. Stiles and A. Zangwill for extensive discussions and comments.  W.M.S.  gratefully acknowledges the support of the Department of Energy through DOE Grant DE-FG02-06ER46278.  

{}


\begin{thebibliography}{}
\bibitem{s-d} The s-d model of Ref.\onlinecite{ZhangLi04} involves two exchange-coupled  magnetizations.  This means that it is in some ways like a ferrimagnet.  In particular, if the two magnetizations have different gyromagnetic ratios and are subject to different anisotropy fields, in addition to a modified Larmor-like mode there will be a second mode at a relatively high frequency, with a small magnetization, driven largely by the exchange field.  We know of no evidence for such a mode.  This implies that the gyromagnetic ratios are similar for the two types of magnetization.
\bibitem{Q-def} We follow the convention of Ref.~\onlinecite{SXZ04}, but one may also write the magnetization flux as $\vec{Q}_{i}\equiv\vec{j}_{i}^{M}$.
\end{thebibliography}
\end{document}